\documentclass[aps,twocolumn,superscriptaddress,groupedaddress,nofootinbib]{revtex4}
\pdfoutput=1

\usepackage{bm}
\usepackage{amsmath,amssymb}
\usepackage{float}
\usepackage{tikz}
\usepackage{tikz-cd}
\newcommand{\Tr}{\text{Tr}} 
\newcommand{\tr}{\text{tr}} 
\numberwithin{equation}{section}

\begin{document}

\def\a{\alpha}
\def\b{\beta}
\def\c{\varepsilon}
\def\d{\delta}
\def\e{\epsilon}
\def\f{\phi}
\def\g{\gamma}
\def\h{\theta}
\def\k{\kappa}
\def\l{\lambda}
\def\m{\mu}
\def\n{\nu}
\def\p{\psi}
\def\q{\partial}
\def\r{\rho}
\def\s{\sigma}
\def\t{\tau}
\def\u{\upsilon}
\def\v{\varphi}
\def\w{\omega}
\def\x{\xi}
\def\y{\eta}
\def\z{\zeta}
\def\D{\Delta}
\def\G{\Gamma}
\def\H{\Theta}
\def\L{\Lambda}
\def\F{\Phi}
\def\P{\Psi}
\def\S{\Sigma}

\def\o{\over}
\newcommand{\gsim}{ \mathop{}_{\textstyle \sim}^{\textstyle >} }
\newcommand{\lsim}{ \mathop{}_{\textstyle \sim}^{\textstyle <} }
\newcommand{\vev}[1]{ \left\langle {#1} \right\rangle }
\newcommand{\bra}[1]{ \langle {#1} | }
\newcommand{\ket}[1]{ | {#1} \rangle }
\newcommand{\EV}{ {\rm eV} }
\newcommand{\KEV}{ {\rm keV} }
\newcommand{\MEV}{ {\rm MeV} }
\newcommand{\GEV}{ {\rm GeV} }
\newcommand{\TEV}{ {\rm TeV} }
\newcommand{\1}{\mbox{1}\hspace{-0.25em}\mbox{l}}
\newcommand{\headline}[1]{\noindent{\bf #1}}
\def\diag{\mathop{\rm diag}\nolimits}
\def\Spin{\mathop{\rm Spin}}
\def\SO{\mathop{\rm SO}}
\def\O{\mathop{\rm O}}
\def\SU{\mathop{\rm SU}}
\def\U{\mathop{\rm U}}
\def\Sp{\mathop{\rm Sp}}
\def\SL{\mathop{\rm SL}}
\def\tr{\mathop{\rm tr}}
\def\mpl{M_{\rm Pl}}

\def\IJMP{Int.~J.~Mod.~Phys. }
\def\MPL{Mod.~Phys.~Lett. }
\def\NP{Nucl.~Phys. }
\def\PL{Phys.~Lett. }
\def\PR{Phys.~Rev. }
\def\PRL{Phys.~Rev.~Lett. }
\def\PTP{Prog.~Theor.~Phys. }
\def\ZP{Z.~Phys. }

\def\dd{\mathrm{d}}
\def\ff{\mathrm{f}}
\def\BH{{\rm BH}}
\def\inf{{\rm inf}}
\def\ev{{\rm evap}}
\def\eq{{\rm eq}}
\def\SM{{\rm sm}}
\def\Mpl{M_{\rm Pl}}
\def\GeV{{\rm GeV}}
\newcommand{\Red}[1]{\textcolor{red}{#1}}
\newcommand{\RC}[1]{\textcolor{blue}{\bf RC: #1}}

\title{
Quark and lepton mass matrices 
	from localization in M-theory on $G_2$ orbifold
}

\author{Eric Gonzalez}
\email{ericgz@umich.edu}
\affiliation{\small Leinweber Center for Theoretical Physics, Department of Physics, University of Michigan, Ann Arbor, Michigan 48109, USA}
\author{Gordon Kane}
\email{gkane@umich.edu}
\affiliation{\small Leinweber Center for Theoretical Physics, Department of Physics, University of Michigan, Ann Arbor, Michigan 48109, USA}
\author{Khoa Dang Nguyen}
\email{kdng@umich.edu}
\affiliation{\small Department of Mathematics, University of Michigan, Ann Arbor, MI 48109, USA }
\author{Malcolm J. Perry}
\email{malcolm@damtp.cam.ac.uk}
\affiliation{\small DAMTP, Cambridge University, Centre for Mathematical Sciences, Wilberforce Road,Cambridge CB3 0WA, UK}
\affiliation{\small Department of Physics, Queen Mary University of London, Mile End Road, Bethnal Green, London E1 4NS, UK.}

\begin{abstract}
	M-theory compactified on a $G_2$ manifold with resolved $E_8$ singularities realizes 4d $\mathcal{N} = 1$ supersymmetric gauge theories coupled to gravity with three families of Standard Model fermions. Beginning with one $E_8$ singularity, three fermion families emerge when $E_8$ is broken by geometric engineering deformations to a smaller subgroup with equal rank. In this paper, we use the local geometry of the theory to explain the origin of the three families and their mass hierarchy. We linearize the blowing-up of 2-cycles associated with resolving $E_8$ singularities. After imposing explicit constraints on the effectively stabilized moduli, we arrive at Yukawa couplings for the quarks and leptons. We fit the high scale Yukawa couplings approximately which results in the quark masses agreeing reasonably well with the observations, implying that the experimental hierarchy of the masses is achievable within this framework. The hierarchy separation of the top quark from the charm and up is a stringy effect, while the spitting of the charm and up also depends on the Higgs sector. The Higgs sector cannot be reduced to having a single vev; all three vevs must be non-zero.Three extra $U(1)$’s survive to the low scale but are not massless, so Z’ states are motivated to occur in the spectrum,  but may be massive.
\end{abstract}
\maketitle

\pagebreak

\section{Introduction}

 M-Theory has been met with considerable success  \cite{Acharya:2001gy,Acharya:2000gb,Acharya:2007rc,Acharya:2008zi}. One prediction of compactified M-Theory is the existence of $\mathcal{N}=1$ supersymmetry and and its soft breaking via gluino condensation, while simultaneously stabilizing all moduli~\cite{Acharya:2008zi, Papadopoulos1995}. M-Theory accommodates radiative electroweak symmetry breaking \cite{Acharya:2007rc}, baryogenesis \cite{Kane:2019nes}, a solution to the strong CP problem \cite{Acharya:2010zx}, and a mechanism for inflation \cite{Kane:2019nod}. Lastly, this framework can include a wide variety of hidden sector dark matter candidates and predict a supergravity spectrum semi-qualitatively \cite{Acharya:2007rc,Acharya:2008zi}. Moreover, most results from string theories can be extrapolated to M-theory through duality. 

In this paper we focus on an M-theory calculation of the quark and charged lepton masses. The first step is to find an appropriate reduction from eleven to four dimensions. Suppose that spacetime is a product $ \mathbb{R}^{3,1} \times X$ where $X$ is a compact 7-d manifold roughly Planck scale in size. Gauge coupling unification and M-theory compactification hint at unbroken supersymmetry at the unification scale. Berger’s theorem \cite{Acharya:2004qe} requires that $\mathcal{N}=1$ SUSY implies that the holonomy group of the manifold $X$ is $G_2$. The resultant low-energy theory can only contain $U(1)$ gauge fields. Such a compactification scheme is unrealistic since the SM contains non-Abelian gauge fields. One introduces singularities into $X$ to ameliorate this issue. A special type of singularity called ADE \footnote{ADE stands for A, D, and E Lie algebra.} allows non-Abelian gauge groups to exist in the theory. Suppose that the local model of $X$ with ADE singularity is of the form $\mathbb{C}^2/\Gamma \times \mathbb{R}^3$, where $\Gamma$ is a finite subgroup of $SU(2)$ (see Table $\ref{diagram}$). Under these circumstances, a super Yang-Mills $\mathcal{N}=1$ multiplet with gauge group $G = SU(k), SO(2k), E_6, E_7$ and $E_8$ respectively will be supported. These singularities can be deformed to break the symmetry of the gauge group $G$ to a subgroup of $G$ with equal rank.


We focus on breaking $\bf 248$ of $E_8$ to SM particle. The matter that survives the symmetry breaking process consists of three multiplets in the {\bf 27} representation of $E_6$, and none in the $\bf{\overline{27}}$ representation of $E_6$ \footnote{ Acharya et al \cite{Bourjaily:2008ji} explained that the net number of chiral zero modes was one. So, either ${\bf 27}$ or ${\bf \overline{27}}$ was
	a normalizable zero mode, but not both. As a convention, we pick the normalizable zero mode to be in ${\bf 27}$. } \cite{Bourjaily:2008ji}. This can explain why there are three and only three families. We explore the aforementioned symmetry breaking pattern by looking to see if a realistic SM theory can descend from a compactified M-theory construction. We calculate the Yukawa couplings under the assumption that everything originates from a deformed $E_8$ theory where the singularity is resolved into a lesser ranked singularity which is associated with $SU(3)\times SU(2) \times U(1)\times U(1)^4$ gauge group \footnote{We separate one $U(1)$ factor out to emphasize SM gauge group.}.

To explain the origin of the three families and their mass hierarchy, breaking $E_8$ to the SM by the traditional Higgs mechanism has been unsuccessful and has shown a lack of predictability, while geometrically engineered M and F theories with $E_8$ points offer an alternative method of symmetry breaking. Moreover, \cite{Bourjaily:2009vf} and related works suggest M-theory based on an $\hat{E}_8-ALE$ space provide more predictability than the analogous model in F-theory\cite{Marchesano:2015dfa,Tatar:2006dc,Beasley:2008kw}. Finally, a description of a singular $G_2$ manifold with Higgs bundles provides a formulation which makes explicit computation of Yukawa couplings possible \cite{Pantev:2009de,Braun:2018vhk}.

We are interested in explicitly calculating the hierarchy of quark mass matrices. As that would include an explicit method for computing matter content, in gauge symmetry breaking through deformation, and their coupling constants, the results would be applicable to a wider study of other matter interaction. We also compute the mass matrix for charged leptons. 

The paper is aimed at a wider audience, so some technical details are omitted and referred to external sources. Section~\ref{intro} contains a brief review of M-theory on a $G_2$ manifold. Section~\ref{singularity} describes the resolution of $ADE$ singularities and the method for computing gauge group symmetry breaking.  In section~\ref{E_8} we explicitly compute this breaking for the $E_8$ singularity with an explicit example of how to compute and locate the fermions on $M_3$. Section \ref{Yukawasection} discusses the general computation for the Yukawa couplings in a local model which leads to explicit quark and lepton terms in section~\ref{quark}. After some gauge fixing for base-space $M_3$'s parameters, numerical results are discussed in section \ref{numerical}. We see that the physical hierarchy is achievable with a very small set of solutions, putting a stringent constraint on the moduli of the theory. Section \ref{effect} discusses the roles of both Yukawa couplings and Higgs  vacuum expectation values (VEVs) in this hierarchy. 
\section{A Brief Background of M-theory on $G_2$ Singular Manifolds}\label{intro}

M-theory is an 11 dimensional theory that can be compactified on a compact 7d manifold $X$ while the remaining non-compact four dimensions are the classical 4 space-time. In the supergravity limit, $X$ is a necessarily a $G_2$ manifold. Moreover, charged chiral particles are only possible on a singular $G_2$ manifold \cite{Acharya:2004qe}. The simplest local model for such 7d manifold is given by the fibering of $\widehat{\mathbb{C}^2/\Gamma_{ADE}}$ over the base $M_3$. Here, $M_3$ is an associative 3-cycle \footnote{Equations of motion requires minimal volume, and an associative cycle is a minimal volume cycle.} in the $G_2$ manifold. $\Gamma_{ADE}$ is a finite subgroup of $SU(2)$ acting on $\mathbb{C}^2$. $\mathbb{C}^2/\Gamma_{ADE}$ is an  asymptotically locally Euclidean manifold (ALE) with ADE singularity at the origin. $\widehat{\mathbb{C}^2/\Gamma_{ADE}}$ denotes any manifold achieved from $\mathbb{C}^2/\Gamma_{ADE}$ by partially smoothing (resolving) the singularity. Locally, the manifold is of the form
\begin{align}
\mathbb{R}^{3,1}\times M_3 \times \widehat{\mathbb{C}^2/\Gamma_{ADE}}
\end{align}
Note that globally, the fiber $\widehat{\mathbb{C}^2/\Gamma_{ADE}}$ varies along the base $M_3$ where the singularity can be smoothed out to different degrees. More details on a recent construction of compact $G_2$ manifolds are in \cite{Kovalev:2003aa,Braun:2017ryx,Braun:2017uku,Corti:2012kd}.

\subsection{Gauge Group Enhancement}\label{field content}
Inherited from supergravity at low-energy limit, the basic fields are a metric $g$, a 3-form potential $C_3$, and a gravitino spinor $\Psi$. We will briefly review the essential properties of the fields needed for this paper. More details are discussed in the appendix  and \cite{Pantev:2009de,Braun:2018vhk,Kennon:2018eqg,Halverson:2015vta}.
From Chern-Simon (CS) terms, $C_3$ is integrated over a manifold of the same dimension, i.e a 3 submanifold of space-time. Excluding time, this submanifold is 2d spatial. This 2d submanifold is an $M_2$ brane. We say $C_3$ electrically couples with $M_2$ brane. Dimensional reduction of the $C_3$ form on the $ALE$ fiber produces $U(1)$ gauge fields
\begin{align}\label{reduction}
C_3 = A_i \wedge \omega^i + \dotsc
\end{align}
where $A_i$'s are one forms (vector fields) on $\mathbb{R}^{3,1}$, and $\omega^i$'s are harmonic two forms associated with 2-cycles of ALE fibers.

The non-abelian gauge group is produced in a similar manner as n coincident D6-branes in type IIA string theory \cite{Sen:1997kz}.
In another perspective independent of duality, the gauge symmetry at a ADE singularity comes from the symmetry of differential form under automorphism of the resolved manifold. Explicitly, the two forms on the resolved manifold can be expressed as element of the lie algebra of the associated ADE group. Therefore, under automorphic map on the resolved manifold, the form can be transformed under the action of the lie group. At singular points where some cycles shrink to a single point, the forms in the same orbit under the transformation induced from the automorphism of those cycles correspond to the same state, so the transformation is a gauge transformation. For example, a self-contained description for the gauge transformation from $SU(N)$  singularity, i.e, $A_{N-1}$ type would be summarized in the below diagrams. The $C_3$ is decomposed into the basis of the 3-forms. In the local description, the basis elements contains components that are 2-forms $\alpha_i$ on the 2-spheres $\mathbb{CP}^1$ which resolves the singularity. 
\[ \begin{tikzcd}
 \text{$G_2$ manifold} \arrow[swap]{dd}{\text{locally}} &&  \text{$C_3$ field} \arrow[swap]{dd}{\text{decompose}}\\ \\
 \text{$M_3 \times \widehat{\mathbb{C}^2 / \Gamma}$} \arrow{d} \arrow{rr}{\text{1-forms $\wedge$ 2-forms }} \arrow[swap]{rr}{\text{basis}} && \text{$\phi_I \wedge$}  \text{$\alpha_J$} \arrow{d}\\
  \text{$\widehat{\mathbb{C}^2 / \Gamma}$} \arrow{dd}{\text{embeded as  $\mathbb{CP}^1$}}     && \text{$\alpha_J$} \arrow{dd}{\text{lifted to}}\\ \\
\text{$\mathbb{C}^N$} \arrow{rr}{\text{2-forms on 2-spheres }} \arrow[swap]{rr}{\text{centers at origins}} && \text{$A_{ij} dz_i  \wedge d\tilde{z}_j$} 
\end{tikzcd} \]
When embedding $\widehat{\mathbb{C}^2 / \Gamma}$ into $\mathbb{C}^N$, we can explicitly write $\alpha_i$ in a local coordinate and see the gauge field $A_{ij}$ transforming under the rotations of $SU(N)$. Fibering this on the $M_3$ base, we see the corresponding adjoint-valued form $\phi$ mentioned in \cite{Braun:2018vhk}
\[ \begin{tikzcd}
\text{Symmetry of $\widehat{\mathbb{C}^2 / \Gamma}$} && \text{$A_{ij} \in \mathfrak{su}(N)$} \arrow[swap]{ddd}{\text{Integrate the 2-cycles}}  \\
 \text{is explicitely rotations $SU(N)$} \arrow{dd}{\text{fibering on}} && \\ \\
\text{$M_3$} \arrow{rr}{\text{Higgs bundle}} && \phi_I \otimes A_J
\end{tikzcd} \]\\	
where $\phi \equiv \sum_{I,J}\phi_I \otimes A_J$ is explicitly an field transform in adjoint of $SU(N)$ (through $A_{ij} \in \mathfrak{su}(N)$), thus befitting the $SU(N)$ gauge description. Similarly, we can embed $D_N$, $E_6$, $E_7$, and $E_8$ type singularities into $\mathbb{R}^{2N}$, $\mathbb{C} \otimes \mathbb{O}$ (bioctonions), $\mathbb{H} \otimes \mathbb{O}$ (quateroctonions), and $\mathbb{O} \otimes \mathbb{O}$ (octooctonions) respectively.

The moral of this is the gauge symmetry comes from the geometrical symmetry of $\widehat{\mathbb{C}^2 / \Gamma}$ which can be explicitly realized by embedding into a covering space. This is an explicit connection to 7d super Yang-Mills theory on $\mathbb{R}^{3,1} \times M_3$ by Higgs bundle.(The connection has been known for a long time through duality without explicit embedding).

It has always been mentioned that $M_2$ branes wrapping ADE singularities will give non-abelian gauge. In here, we can see gauge boson $A_{ij}$ explicitly and independently from the duality description. 

In a more intuitive sense, the warping of $M_2$ branes around non-vanishing ALE cycles creates massive vector bosons. The masses are proportional to the volume of the 2-cycles. By shrinking the 2-cycles, we are making those massive bosons massless, Moreover, the configuration of the 2-cycles (Dynkin diagram) dictates the relation of these bosons and fits them perfectly into an non-abelian gauge group. Inversely, at any point on $M_3$ where the volume of a 2-cycle is non-zero, the associated vector boson becomes massive and hence must be removed from the gauge group. Yet, the $U(1)$ in the Cartan subalgebra from (\ref{reduction}) is unaffected by this, so we still have a $U(1)$ gauge symmetry. Hence, the n-ranked gauge group is broken into an $(n-1)$-ranked subgroup and a $U(1)$ (total rank is unchanged). In general, each non-vanishing volume of a basis 2-cycle reduces the rank of the group by one and leave a $U(1)$ behind. It is important to note that this is similar to the Higgs mechanism except that the Higgsing happens due to the geometry instead of the traditional Higgs doublets as we will discuss in the next section.
\subsection{Chiral Fermion}
On a singularity curve for a non-abliean gauge group $H$, which is a resolution \footnote{See section \ref{singularity}.} of higher rank singularity of a larger gauge group $G$, chiral fermion solutions are localized at points where the singularity associated with $H$ is worsened by a conical singularity \cite{Acharya:2001gy,Atiyah:2001qf,Bourjaily:2008ji,Berglund:2002hw}. By considering the resulting extra subgroup generated by the extra shrunk two cycles, one can determine the representation of the fermions with respect to the gauge group $H$. We will elaborate this in $\ref{fermion_rep}$.
\section{ADE Singlarity, Resolution, and Deformation}\label{singularity}

ADE singularity classifies a family of singularities that has an injective map into the set of general unbroken gauge symmetries in M-theory as a consequence of the Mckay correspondence. Therefore, we briefly review ADE singularity classification \footnote{Originally due to \cite{Duval:1934aa}. ALE construction by hyper-K\"ahler quotients is in \cite{Kronheimer:1989zs}.}. An ADE singularity can be written as $\mathbb{C}^2/\Gamma$ where $\Gamma \subset SU(2)$ is a finite subgroup and acts on $\mathbb{C}^2$ by ordinary multiplication. This action has no fixed point other than the origin. Consequently, $\mathbb{C}^2/\Gamma$ has a singularity at the origin .

\begin{figure}
	\includegraphics[scale=0.4]{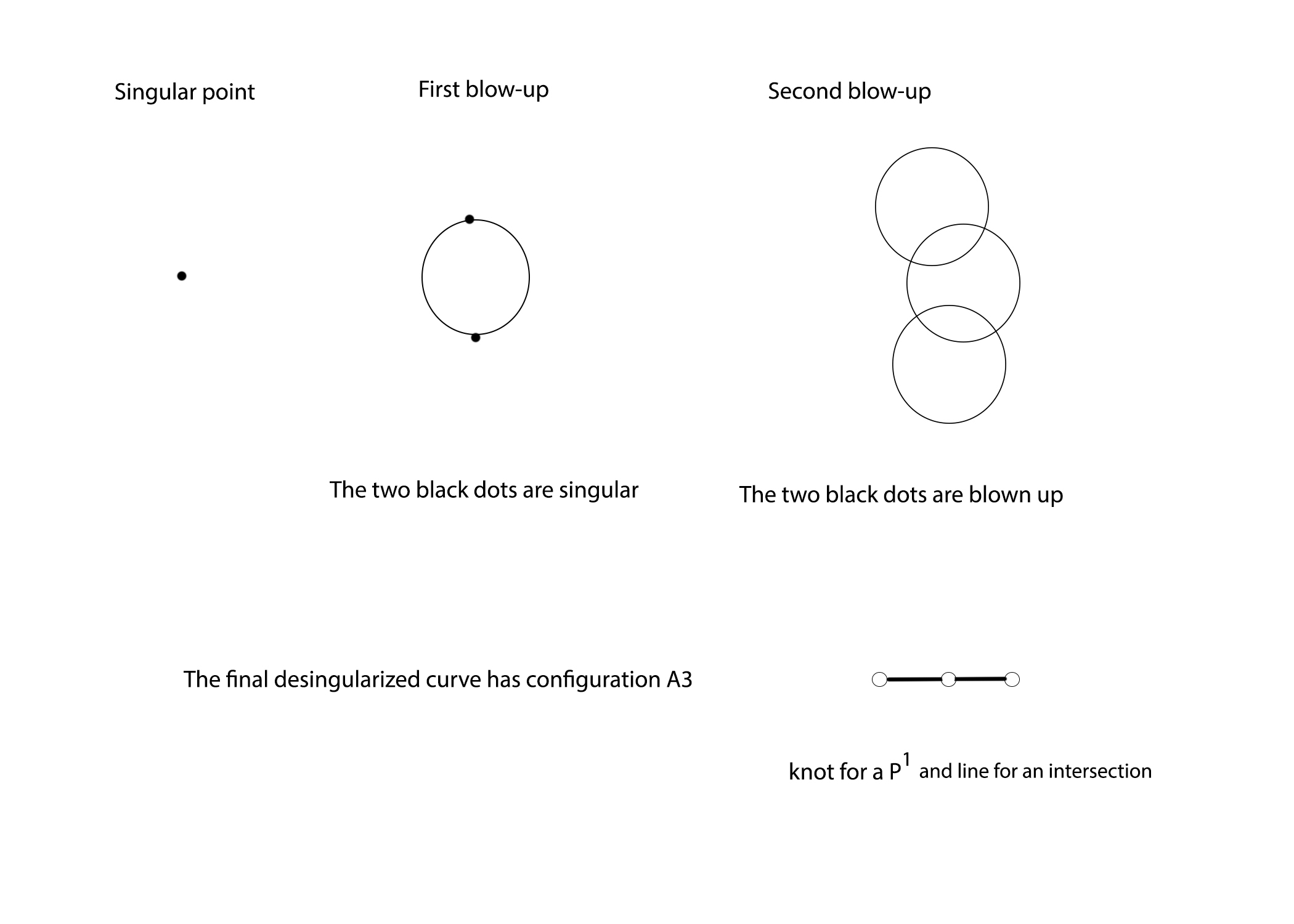}
	\caption{An $A_3$ type singularity being fully resolved will have a configuration of three Riemann spheres $\mathbb{P}^1\mathbb{(C)}$ which intersect according to $A_3$ Dynkin diagram. Note that every two spheres intersect at most at one point transversely.}
	\label{blow_up}
\end{figure}

\begin{figure}[H]
	\includegraphics[scale=0.2]{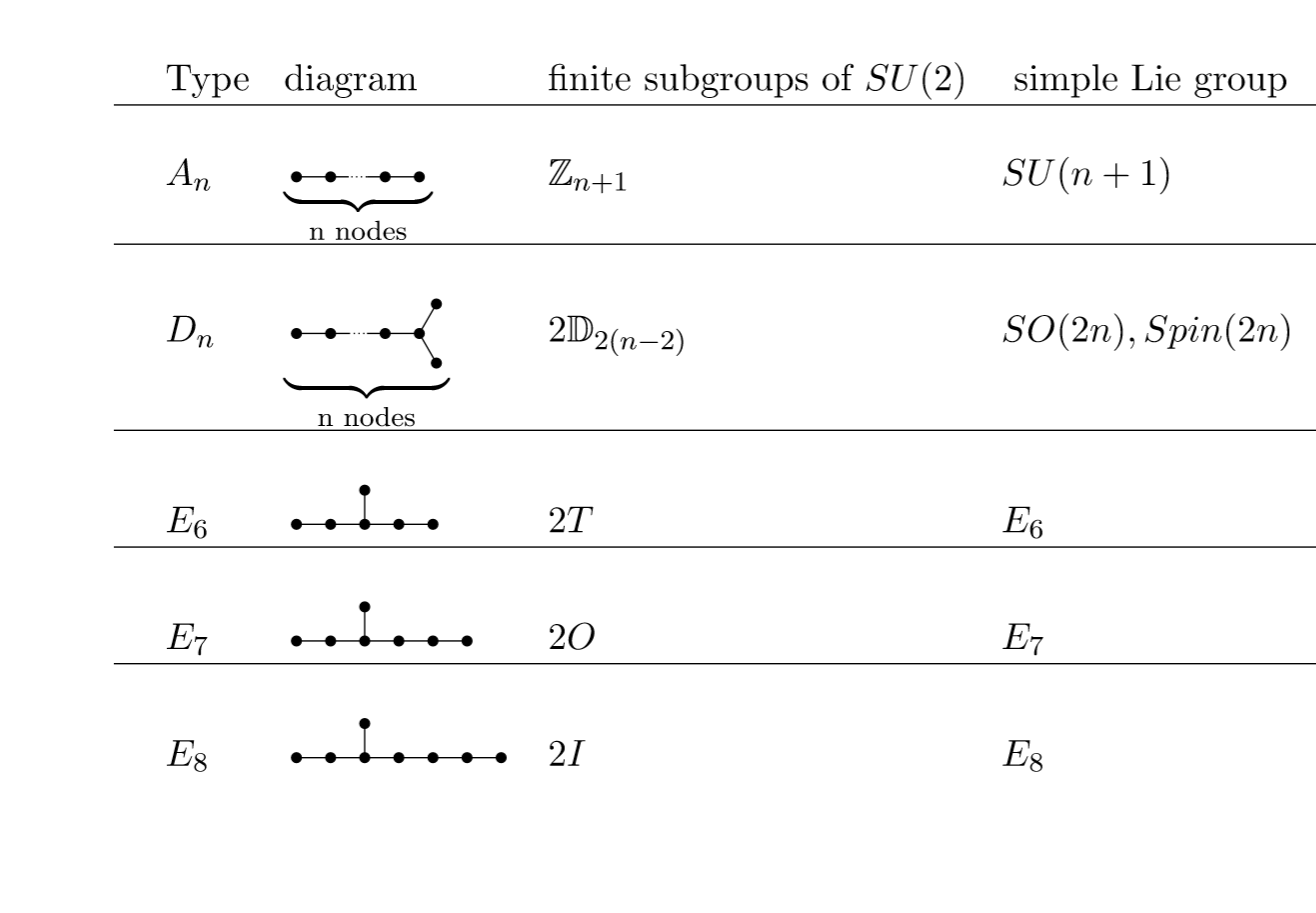}\nonumber
	\caption{Dynkin diagram and associated groups.}
	\label{diagram}
\end{figure}

Such a singularity can be made smooth by expanding the singular point into a projective space $\mathbb{P}^1(\mathbb{C})$ (topologically just a 2-sphere). This procedure is called ``blowing-up", and the blown-up space is called a resolution of the original space. However, the blown-up point may not be resolved completely and still have some remaining singular points on the $\mathbb{P}^1$. We have to keep blowing up those points until there is no singularity. The result is a collection of $\mathbb{P}^1$'s intersecting of each other. The intersection pattern is exactly the Dynkin diagram of the type of singularity. Figure \ref{blow_up} gives a  pictorial illustration of a singularity of type $A_3$.
Each of the consequent $\mathbb{P}^1$ can be called a two-cycle. So, a singularity of type $A_3$ is one that, when completely resolved, has a configuration of $A_3$. Similarly, a singularity of a certain Dynkin diagram has the blown-up configuration of that diagram. The explicit diagrams with the associated group are in Figure \ref{diagram}.

We have seen that the 2-cycles $\mathbb{P}^1$ directly relate to the smoothing of singularities. We can use the volume of the 2-cycles to parametrize the resolution. Such a method of smoothly parametrizing the blowing-up is called deformation. 

For each 2-cycle, we use a harmonic one-form $\phi$ \footnote{This is in fact the VEV of the Higgs field $\phi$ in 7d Yang-Mills theory we mentioned earlier.} on $M_3$, which can be thought of as a metric-invariant 3-vector field on $M_3$, to parametrize the size of the 2-cycle.  Alternatively, Katz et al \cite{Bourjaily:2009vf,Katz:1996xe} use the coefficients in the Cartan subalgebra as the parameters. Consistently, there is a one-to-one bijection between the two parametrizations given by Table \ref{rootvolume}. Following the existing literature,  we denote $\widehat{G}(f_1,f_2, f_3, \dots,f_n)$ as the family of $\widehat{\mathbb{C}^2/\Gamma_{G}}$ parametrized by the coordinates $f_i$ in Cartan subalgebra where $n$ is the rank of $G$ and use Table \ref{rootvolume} to compute the ``volume" one-form $\phi$ when needed \footnote{More details on root system and deformation are in \cite{Katz:1992aa}.}.

\section{ $E_8$ Breaking}\label{E_8}

Our goal is to describe all the particles by resolving one single ADE singularity. $E_8$ is the only simple Lie group that does the job. $E_8$ and its breaking have been studied by several authors \cite{Marchesano:2015dfa,Clemens:2019flx,Bourjaily:2007vx,Pantev:2009de,Dudas:2009hu,Bizet:2014uua,Godazgar:2013rja,Evslin:2003kn,Palti:2012aa}. To understand the breaking, we first explicitly write down the simple roots of $E_8$ in the Dynkin diagram order (see Table \ref{diagram}) 
where $e_i$'s are orthogonal vectors in $\mathbb{R}^{n,1}$. Let $\widehat{E}_8 (f_1, ...,f_8)$ be the resolution of a $E_8$ singularity parametrized by deformation moduli $f_i$'s which are one-forms on $M_3$. The simple roots are associated with the volumes of the blown-up 2-cycles by Table \ref{rootvolume} \cite{Bourjaily:2009vf}.

\begin{table}
    \centering
	\scalebox{0.7}{\begin{tabular}{l  c | c}
	\hline
	 &Positive Roots of $E_n$ & Volume of Corresponding Two-Cycle\\
	\hline
	 & $e_i-e_{j>i}$ & $f_i-f_{j>i}$\\
	 &$-e_0 + e_i + e_j +e_k$ & $f_i +f_j + f_k$\\
	$n \ge 6$ & $-2e_0 + \Sigma^6_{j=1} e_j$ & $\Sigma^6_{j=1} f_j$\\
    n=8 & $-3e_0 + e_i + \Sigma^8_{j=1} e_j$ & $ f_i + \Sigma^8_{j=1} f_j$\\
	\hline
	\end{tabular}}
	\caption{Positive roots of $E_n$ and the associated one-forms (sometimes called ``area" in literature) controlling the sizes of 2-cycles on the ALE fiber. This is Table 1 in \cite{Bourjaily:2009vf} with permission. }
	\label{rootvolume}
\end{table}

\begin{figure}
    \centering
    \includegraphics[scale=0.5]{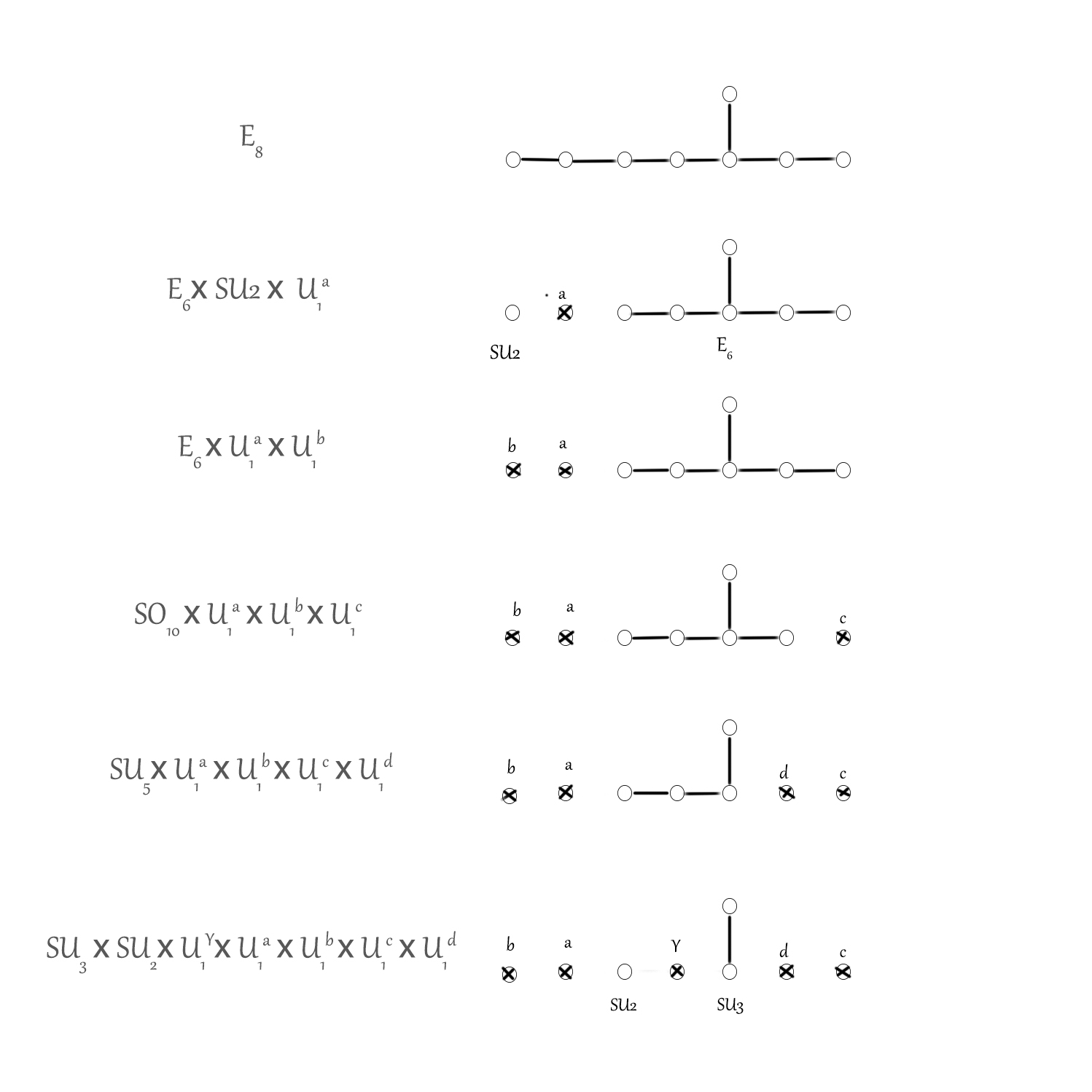}
    \caption{Breaking of $E_8$ by resolving singularity}
    \label{breakingE8}
\end{figure}

Each simple root, or equivalently each knot on the Dynkin diagram, will initially represent a vanishing cycle at the singularity. To break a group to a smaller group, we will ``cut" a knot on their diagram so that we get the diagram of the smaller group. Each ``cutting" is performed by blowing up the cycle (which was initially vanishing) associated with the knot. We recall that each cycle in the above Dynkin diagram gives rise to a boson whose mass is proportional to the volume of the cycle. Therefore, a vanishing cycle in the above Dynkin diagram will result in a massless boson. The goal is to keep the SM gauge bosons massless (zero volume cycles) while the other bosons are massive (non-zero volumn cycles). We will follow the breaking path \footnote{Different paths to the same subgroup will lead to the same physics. This is because if there is a diffeomorphism between $X_1$ and $X_2$ so that their hyper-K\"ahler structures agree, then they are isometric.} of \cite{Bourjaily:2007vx}. Figure \ref{breakingE8} summarizes the above steps. In the figure, we start with an $E_8$ singularity which corresponds to $\hat{E}_8 (0,0,0,0,0,0,0,0)$, then turn on the volumes of the cycles associated with the crossed knots by giving non-zero values for one-form $f_i$'s. There are five volumes needed to be turned on, so we parameterize  $f_i$'s by five non-zero one-forms $a,b,c,d$ and $Y$ (note that $Y$ here is the one-form associated with hypercharge $U(1)^Y$, not the hypercharge itself). They are simply parameters that are linearly combined in a specific way so that the volumes of the cycles vanish or  blow up appropriately by Table \ref{rootvolume}. Then the final manifold is parameterized as \cite{Bourjaily:2009vf}

\begin{align}
\widehat{E}_8 (a+b+c+d+\frac{2}{3}Y, a-b+c+d+\frac{2}{3}Y,\\
-c-d-\frac{7}{3}Y, -c-d-\frac{7}{3}Y, -c-d + \frac{8}{3}Y, \nonumber \\
-c-d + \frac{8}{3}Y, -c + 3d- \frac{4}{3}Y, 2c-2d -\frac{4}{3}Y).\label{resolvedE8}
\end{align}
We can check each step of Figure \ref{breakingE8} by setting all $a,b,c,d,$ and $Y$  in (\ref{resolvedE8}) to zero, then turn them on accordingly to each step, and compute the volumes using Table \ref{rootvolume}. In the following, we can check the volumes of the cycles corresponding to the simple roots in the final step
\begin{equation}
	\begin{pmatrix}\label{cyclevolume}
e_1 - e_2 & 2b\\
e_2 - e_3 & a - b + 2c + 2d +3Y \\
e_3 - e_4 &  0\\
e_4 - e_5 & -5Y \\
e_5 - e_6 & 0 \\
e_6 - e_7 & -4d +4 Y \\
e_7 - e_8 & -3c + 5d\\
- e_0 + e_6 +e_7 + e_8& 0	
\end{pmatrix} 
\end{equation}
This is exactly the configuration of Figure \ref{breakingE8}. Note that one can use any different set of one-forms as long as they fulfill the desired configuration and sufficiently parameterize the independent non-vanishing cycles. \\
Therefore, whatever constrain we make, to avoid an unwanted shrunk cycle which will lead to an extra massless boson, we have to make non-zero volumes in the above table remain non-zero. The would mean
\begin{align}\label{cyclevolume}
	  b &\neq 0 &  a -b + 2c +2d + 3Y &\neq 0&&\\
  Y &\neq 0 &  Y &\neq d &c &\neq \frac{5}{3} d.
\end{align}
\subsection{Fermion Representations}\label{fermion_rep}
Given a gauge group $H$ for the theory, the corresponding cycles on the fiber are shrunk everywhere along the base manifold $M_3$. Those cycles correspond to the simple roots of $H$. A matter representation happens at the points where additional cycles associated with positive roots (see Table \ref{rootvolume})  vanish. By letting the positive roots vanish one by one, we can find all the resulting representations. We will do a few examples showing how to calculate the representation.\\
First, we consider $e_2-e_3$ cycle. Using the above table, we conclude that the associated volume is $f_2 - f_3 = a-b+2c +2d+3Y$. Now, we consider the curve where this particular cycle vanishes: $ a-b+2c +2d+3Y=0$. In order to know what representation emerges at this curve, we consider what kind of weight diagram is generated from $e_2-e_3$ and the roots from the gauge group (corresponding to the globally shrunk cycles) $e_3-e_4$ (corresponding to $SU(2)$), and $e_5-e_6$ and $-e_0+e_6+e_7+e_8$ (corresponding to $SU(3)$). In more details, we will try to find what are the positive roots we can get from $e_2-e_3$ by adding or subtracting $e_3-e_4$ ,  $e_5-e_6$, and $-e_0+e_6+e_7+e_8$. \\
\begin{tikzcd}
&SU(2) &SU(3)\\
& e_2-e_4 & (\text{No positive root}\\
& & \text{from adding or substracting}  \\
&   &e_5-e_6 \text{ or } \\
& &-e_0+e_6 + e_7 + e_8)\\
& e_2-e_3  \arrow[uuuu, swap, "e_3-e_4"] & e_2-e_3\\
\end{tikzcd}\\
From above, we see that there are two positive roots corresponding to $SU(2)$, so the particle will behave like $\pmb{2}$ of $SU(2)$. Only one positive root for $SU(3)$ case, so it is a singlet for $SU(3)$. Thus, this is a $(\pmb{2},\pmb{1})$ of $SU(2)\times SU(3)$ (corresponding to $H_2^u$ as in the Table \ref{matter}). Notice that above calculation implies that $e_2-e_4$ yields the same particle.\\
Next, let's try another positive root, say $-e_0+e_2+e_3+e_5$. The curve equation is $f_2 + f_3 + f_5 = a-b-c-d+Y=0$. Then, we get\\
\begin{tikzcd}	
&SU(2) &SU(3)\\
& -e_0+e_2+e_3+e_5 & -2e_0 + +e_2+e_3\\
& &+e_5 +e_6+e_7+e_8\\
& -e_0+e_2 +e_4 +e_5  \arrow[uu, swap, "e_3-e_4"] & -e_0+e_2+e_3+e_5  \arrow[u, swap, "-e_0 + e_6+e_7 +e_8"]\\
& & -e_0+e_2+e_3+e_6  \arrow[u, swap, "e_5-e_6"]\\
\end{tikzcd}
So by counting the positive roots, we conclude that it is $\pmb{2}$ for $SU(2)$ and $\pmb{3}$ or $\pmb{\bar{3}}$ for $SU(3)$. As fundamental and anti-fundamental are just a convention, we call this order of adding $e_5-e_6$ and $-e_0+e_6+e_7+e_8$ associated with fundamental $\pmb{3}$. Thus this is a  $(\pmb{2},\pmb{3})$ of $SU(2)\times SU(3)$.\\
Lastly, for completeness, we will illustrate the case of $\pmb{\bar{3}}$ with $-e_0+e_2+e_3+e_4$. The curve equation is $f_2 +f_3 +f_4 = a-b-c-d-4Y=0$. Then, we get
\begin{tikzcd}	
&SU(2) &SU(3)\\
&\text{No other positive root} & -2e_0+e_2+e_3 \\
& &+e_4 +e_5+e_7+e_8\\
& -e_0 + e_2 +e_3+e_4 & -2e_0+e_2+e_3+e_4  \arrow[u, "e_5-e_6"]\\
& &+e_6+e_7+e_8 \\
& & -e_0+e_2+e_3+e_4  \arrow[u, "-e_0 + e_6+e_7 +e_8"]\\
\end{tikzcd}
Notice that the order of adding  $e_5-e_6$ and $-e_0+e_6+e_7+e_8$ is reversed from the previous case, so, by above convention, this is a $(\pmb{1},\pmb{\bar{3}})$ of $SU(2)\times SU(3)$.\\
Bourjaily et al \cite{Bourjaily:2009vf} have already worked out the breaking for us. The charges for relevant particles in this paper is presented in Table \ref{matter}. The location of the singularity associating with a particle is a linear combination of moduli weighted by the charges. For instance, the location of $Q_1$ is the curve that satisfies  
\begin{align}\label{sample}
     a+b-c-d+Y=0
\end{align}

\begin{table}
	\centering
	\begin{ruledtabular}
	\begin{tabular}{c c c c c c c c}
    & $SU_3$ & $ SU_2$ & $U_1^a$ & $U_1^b$ & $U_1^c$ & $U_1^d$ & $U_1^Y$  \\
    \hline
    $Q_1$ & $\mathbf{3}$ & $\mathbf{2}$ & $1$ & $1$ & $-1$ & $-1$ & $1$ \\
    $Q_2$ & $\mathbf{3}$ & $\mathbf{2}$ & $1$ & $-1$ & $-1$ & $-1$ & $1$ \\
    $Q_3$ & $\mathbf{3}$ & $\mathbf{2}$ & $-2$ & $0$ & $-1$ & $-1$ & $1$ \\
    
    $u_1^c$ & $\mathbf{\bar{3}}$ & $\mathbf{1}$ & $1$ & $1$ & $-1$ & $-1$ & $-4$ \\
    $u_2^c$ & $\mathbf{\bar{3}}$ & $\mathbf{1}$ & $1$ & $-1$ & $-1$ & $-1$ & $-4$ \\
    $u_3^c$ & $\mathbf{\bar{3}}$ & $\mathbf{1}$ & $-2$ & $0$ & $-1$ & $-1$ & $-4$ \\
    
    $d_1^c$ & $\mathbf{\bar{3}}$ & $\mathbf{1}$ & $1$ & $1$ & $-1$ & $3$ & $2$ \\
    $d_2^c$ & $\mathbf{\bar{3}}$ & $\mathbf{1}$ & $1$ & $-1$ & $-1$ & $3$ & $2$ \\
    $d_3^c$ & $\mathbf{\bar{3}}$ & $\mathbf{1}$ & $-2$ & $0$ & $-1$ & $3$ & $2$ \\    
    
    $L_1$ & $\mathbf{1}$ & $\mathbf{2}$ & $1$  & $1$ & $-1$ & $3$ & $-3$ \\
    $L_2$ & $\mathbf{1}$ & $\mathbf{2}$ & $1$ & $-1$ & $-1$ & $3$ & $-3$ \\
    $L_3$ & $\mathbf{1}$ & $\mathbf{2}$ & $-2$ & $0$ & $-1$ & $3$ & $-3$ \\
    
    $H_1^u$ & $\mathbf{\bar{1}}$ & $\mathbf{2}$ & $1$ & $1$ & $2$ & $2$ & $3$ \\
    $H_2^u$ & $\mathbf{\bar{1}}$ & $\mathbf{2}$ & $1$ & $-1$ & $2$ & $2$ & $3$ \\
    $H_3^u$ & $\mathbf{\bar{1}}$ & $\mathbf{2}$ & $-2$ & $0$ & $2$ & $2$ & $3$ \\
        
    $H_1^d$ & $\mathbf{\bar{1}}$ & $\mathbf{2}$ & $1$ & $1$ & $2$ & $-2$ & $-3$ \\
    $H_2^d$ & $\mathbf{\bar{1}}$ & $\mathbf{2}$ & $1$ & $-1$ & $2$ & $-2$ & $-3$ \\
    $H_3^d$ & $\mathbf{\bar{1}}$ & $\mathbf{2}$ & $-2$ & $0$ & $2$ & $-2$ & $-3$ \\
    
    $e_1^c$ & $\mathbf{1}$ & $\mathbf{1}$ & $1$ & $1$ & $-1$ & $-1$ & $6$ \\
    $e_2^c$ & $\mathbf{1}$ & $\mathbf{1}$ & $1$ & $-1$ & $-1$ & $-1$ & $6$ \\
    $e_3^c$ & $\mathbf{1}$ & $\mathbf{1}$ & $-2$ & $0$ & $-1$ & $-1$ & $6$ \\
	\end{tabular}
	\caption{Relevant particles from three families of $E_6$, for a complete listing see \cite{Bourjaily:2009vf}.}
	\label{matter}
	\end{ruledtabular}
	
\end{table}


\section{Yukawa Coupling from Volume of the Three-Cycle}\label{Yukawasection}
In the superpotential, a cubic term $ABC$ is allowed at tree level if the product transforms as a singlet under the gauge group. In particular, that implies the sum of charges for each of the $U(1)$'s is zero. If such a term happens, each of the particles $A, B,$ and $C$ will live on a different conical singularity which corresponds to different points $t_A, t_B,$ and $t_C$ on the base $W$ which are solutions of equations derived from Table \ref{matter} (similar to \ref{sample}). The idea of this section is that the Yukawa coupling coefficient of this term is proportional to the exponential of the volume of the three-cycle wrapping around the three singularities 
\begin{align}
\text{Yukawa coupling}= n_{ABC}\frac{e^{-Vol( \Sigma_{ABC})}}{\Lambda_{ABC}} \label{Yukawa}
\end{align}
where $\Sigma_{ABC}$ is the three-cycle wrapping around the singularities, $n_{ABC}$ is the sign of the term which depends subtly on the orientation of the three cycle\cite{Braun:2018vhk} \footnote{Details of how to determine $n_{ABC}$ is in \cite{Braun:2018vhk} and Appendix F of \cite{Gaiotto:2015aoa}}, $\Lambda_{ABC}$ is a scale factor which is approximately the volume of $G_2$ manifold. We will temporarily ignore both of $n_{ABC}$ and  $\Lambda_{ABC}$ in our analysis in this section. 

We are interested in the limit where gravity decouples. The $G_2$ manifold here is treated as large enough to make the calculation manageable. Then, we can focus on a local patch of $M_3$ which is approximately $\mathbb{R}^3$. The volume of the three-cycle in the linearization has been roughly formulated by \cite{Bourjaily:2009vf}. However, a more complete analysis shows the requirement of the harmonic condition and relative rotations of the fields. By BPS equations \cite{Braun:2018vhk}, locally for each moduli $\phi$ ( $\phi = a, b, c, d,$ and $Y$. These are the $f_i$'s in the previous sections), there is a harmonic function $h_{\phi}$ on  $M_3$ base so that $\phi = d h_{\phi}$ \cite{Braun:2018vhk}. For simplicity, we think of $ \phi$ as a three vector, and  $\phi = \nabla h_{\phi}$ . Harmonic condition requires that $\Delta h_{\phi} =0$. That means 
\begin{align}
\partial_i \phi^i = 0.
\end{align}

This requires that on linear level,
\begin{align}
\phi = H t + v \label{para}
\end{align}
where $H$ is a real traceless symmetric 3x3 matrix, $v$  is a real three vector, $t$ is a local real parametrization of the 3d base. Then, $h_{\phi}$ will have the form
\begin{align}\label{h}
\frac{1}{2} t^T H t + v^T t + c
\end{align}
where c is a constant term. 

The location of a particle, say $X$, is a zero $t_X$ of a linear combination $\phi_X$ of $a,b,c,d$, and $Y$ with by the charges from table \ref{matter}. From previous discussion, $t_X$ is the critical point of a harmonic function $h_{\phi_X}$.  Assume the critical points are isolated. This is the same as assuming $H_{X}$ is invertible. The critical point of $h_{\phi_X}$ or the zero point of $\phi_X$ is
\begin{align}
t_{X}=-H_{X}^{-1}v_{X}.
\end{align}
Then, if the $ABC$ term is allowed, i.e, $h_{\phi_A} + h_{\phi_B} + h_{\phi_C} =0$, the volume for the three-cycle wrapping the three critical points $t_{A}$, $t_B$ and $t_C$ is \footnote{\cite{Braun:2018vhk} gives formulation for the general case, which has been applied to this linear case.}
\begin{align}\label{volume}
Vol(\Sigma_{ABC})&=h_{\phi_A}(t_A) + h_{\phi_B}(t_B) + h_{\phi_C}(t_C) \nonumber \\
&= \frac{1}{2} (-v_A^T H_A^{-1} v_A - v_B^T H_B^{-1} v_B \\
&+ (v_A + v_B)^T (H_A + H_B)^{-1} (v_A+ v_B)).\nonumber
\end{align}
Notice that the constant $c$ in equation (\ref{h}) plays no role here due to cancellation, so in practice, we will simply drop it. In section \ref{Yukawa_compute}, explicit computation for a Yukawa coupling is shown for a quark term.
\subsection{Discussion of Other Features}\label{extra}
So far, we have only considered $M_3$ as a flat $\mathbb{R}^3$ which obviously overlooks the very stringent global structure of a compact $G_2$ manifold. This structure may reduce the parametrization freedom we have in the flat local case. The singularities curves may also cut each other at some point beyond the local area due to compactness, increasing the number of possible Yukawa couplings. Additionally, the sign factors in equation (\ref{Yukawa}) may also change the mass matrix significantly. They are determined by the gradient flow of the $h_\phi$ \cite{Braun:2018vhk,Harvey:1999as,Beasley:2003fx}. It is difficult to study the gradient flow between singular points for the local model as the space is not compact. Future study of the gradient flows and hence the sign factors can reveal more of the mass matrix.\\
As mentioned in section \ref{fermion_rep}, we should project out particles we do not plan to include in our theory. Projecting a specific particle includes requiring that the curves never satisfy the particle's equation  derived from Table \ref{matter}. That would create more restraint on the parameters. For our local case in particular, it would require a vanishing determinant of a certain linear combination of $H_\phi$'s.
Nonetheless, the problems with these particles are not detrimental and can be remedied by other means. Careful study is needed on this issue.
\section{Quark Terms}\label{quark}
\subsection{General Quark Terms}
Recall that the quarks get mass when the Higgses receive VEVs. For example,
\begin{equation}
\lambda^{ij} H^u_k  Q_i u_j \rightarrow	\langle H^u_k \rangle \lambda^{ij} Q^k_i u^k_j.
\end{equation}
Ellis et al \cite{Ellis:2014kla} showed that $\tan \beta \approx 7$, from electroweak symmetry breaking, so we know both up and down VEVs in the two-Higgs-doublets model. We will discuss later how to adapt these into the six Higgs doublets in this paper. Quark terms that satisfy vanishing sum of charges are 
\begin{align}
Q_1 u^c_2 H^u_3 + Q_2 u^c_1 H^u_3 + Q_1 d^c_2 H^u_3 + Q_2 u^c_1 H^u_3 + \nonumber\\
Q_2 u^c_3 H^u_1 + Q_3 u^c_2 H^u_1 + Q_2 d^c_3 H^u_1 + Q_3 u^c_2 H^u_1 + \label{quarksuper}\\
Q_3 u^c_1 H^u_2 + Q_1 u^c_3 H^u_2 + Q_3 d^c_1 H^u_2 + Q_1 u^c_3 H^u_2. \nonumber
\end{align}
Note that there is no diagonal term in this general setting. Also, some couplings between the Higgs and the quarks which could have been possible in SM are forbidden here due to the extra $U(1)$'s. Nonetheless, those terms can still be generated by Giudice-Masiero mechanism after the breaking of supergravity \cite{Casas:1992mk,Acharya:2016kep}. However, we will leave this mechanism to future study in the context of M-theory with $E_8$ orbifold. In the following sections, we will focus on the simplest constraints on the moduli to make the theory physical.

The relevant terms for leptons are
\begin{align}
L_1 e^c_2 H^d_3 + L_2 e^c_3 H^d_1 + L_3 e^c_1 H^d_2 +\\
L_1 \nu^c_2 H^u_3 + L_2 \nu^c_3 H^u_1 + L_3 \nu^c_1 H^\nu_2 .
\end{align}
Notice that we only have Dirac mass terms here. Majorana terms may require quartic level, extra particles getting a VEV, or extra constraints on the moduli, so we will not discuss such terms in this paper.
\subsection{Diagonal Terms and Setting $a=0$}
(\ref{quarksuper}) shows that there is no diagonal term for the quark matrices. This appears to be a problem because with the top quark mass much larger than those of up and charm quarks, the trace of the mass matrix must be non-zero. This problem is generic in our method of constructing three families from $E_8$ singularity. The same issue was discussed in the F-theory context in \cite{Beasley:2008kw}. The reason for this is the conservation of charge in $a$ and $b$. Hence, this directly relates to the separation of families because $a$ and $b$ break the adjoint of $E_8$ into three ${\bf 27}'s$ in $E_6$. So, particles in the same family must have the same charge in $a$ and $b$, making it impossible for them to form a singlet cubic term within the same family in generic setting. One way to remedy this is to introduce a self intersecting curve for the up-type when $Y=0$ \cite{Beasley:2008kw}, using the fact that in grand unified theories $u$ and $Q$ both stay on the same curve of $\bf{10}$ of $SU(5)$. However, this method cannot be applied for down-type as $d$ does not stay on the same curve as $Q$. Moreover, self-intersecting requires higher order then linearization which we will not pursue here. Alternatively, Bourjaily et al \cite{Bourjaily:2009vf} also discuss the contribution of quartic terms. This will require giving large VEVs for extra particles, creating more parameters which we will not consider at this time.\\
In this paper, we can consider some constraint on $a$ and $b$ leading to possible non-zero diagonal terms. This in essence sets a relation for $a$ and $b$ charges. We still keep in mind the condition of non-vanishing volumes in (\ref{cyclevolume}) as we do not wish to unnecessarily enhance the gauge symmetry. The simplest constraint we can make is $a=0$. Although it is intriguing to study other constraints, we will ignore them in this paper. This constraint will restrict the gauge group to  $SU(3)\times SU(2) \times U(1)^Y \times U(1)^b \times U(1)^c  \times U(1)^d$. In term of geometry, this breaking of $U(1)^a$ is equivalent to restricting the basis 2-cycles in a linear relation, reducing the number of independent 2-cycles and hence number of $U(1)$'s.
\subsection{Quark Mass Matrices}\label{Yukawa_compute}
After setting $a=0$ together with the localization, the up-type quark mass matrix can be computed. We will show one example of the computation here for $M_{12}^u u_1 u_2^c$. It comes from the term
\begin{align}
\lambda_{123}^u Q_1 u_2^c H^u_3. \label{Yukawa_sample}
\end{align}
When the Higgs gets VEV at low scale, the term becomes
\begin{align}
\lambda_{123}^u \langle H^u_3 \rangle u_1 u_2^c,
\end{align}
where $M^u_{12}=\lambda_{123}^u \langle H^u_3 \rangle$. Then, all that is left is to compute  $\lambda_{123}^u$. At high scale, $\lambda_{123}^u$ can be calculated from (\ref{volume}) and Table \ref{matter}. In the linearization language
  \begin{align}
H_{Q_1} &=  H_b-H_d +H_Y\\
v_{Q_1} &=  v_b - v_d + v_Y\\
H_{u_2} &= -H_{b} -H_d-4H_{Y}\\
v_{u_2} &= -v_{b} - v_d -4v_{Y}
\end{align}
 then (\ref{volume}) gives 
\begin{align}
&Vol\{\Sigma_{Q_1 u^c_2 H^u_3} \}=\\
&\frac{1}{2} \Big( (v_b  -v_d+v_Y)^T ( H_b  -H_d +H_Y)^{-1}( v_b -v_d+v_Y) +  \nonumber\\
&( - v_b -v_d-4v_Y)^T (- H_b  -H_d - 4H_Y)^{-1}( - v_b -v_d - 4v_Y) +\\
&( 2v_d+3v_Y)^T ( +2H_d +3H_Y)^{-1}(2v_d+3v_Y) \Big) \nonumber
\end{align}   
Thus, ($\ref{Yukawa}$) , ignoring the overall scaling, gives
\begin{flalign}
&\lambda^u_{123} = n^u_{12}  \exp \Big\{ \nonumber\\ 
&-\frac{1}{2} |  (v_b  -v_d+v_Y)^T ( H_b  -H_d +H_Y)^{-1} 
( v_b -v_d+v_Y) +  \nonumber\\
&( - v_b -v_d-4v_Y)^T (- H_b  -H_d - 4H_Y)^{-1}( - v_b -v_d - 4v_Y) + \nonumber\\
&( 2v_d+3v_Y)^T ( +2H_d +3H_Y)^{-1}(2v_d+3v_Y)|  \Big\}
\end{flalign}
Then, we have to run these Yukawa couplings down to the SM scale to compute the mass. Note that for the diagonal term $Q_3 u^c_3 H_3^c$, obtained from setting $a=0$, can be computed by the above method. 
\subsection{Six Higgs VEVs}
In the six Higgs doublets model without extra $U(1)$'s, one can choose a basis for up-type and down-type Higgses so that only one pair of Higgses gets a VEV without loss of generality. Here, due to different charges for the Higgses from the extra $U(1)$'s (see Table \ref{matter}), we cannot make such a choice of basis. 

We will try to translate from  the two VEVs of SM Higgses to the six VEVs in our theory. By standard QFT, we can relate this by looking at the mass of W boson in the SM and identify
\begin{align}
\langle H_u^{SM} \rangle ^2 = \sum_{i}\langle H_u^{i} \rangle ^2,\\
\langle H_d^{SM} \rangle ^2 =  \sum_{i}\langle H_d^{i} \rangle ^2.
\end{align}
So, we can use spherical parametrization to write
\begin{align}\label{spherical}
\langle H_{u/d}^{1} \rangle  &= \langle H_{u/d}^{SM}\rangle \cos{\phi_{u/d}}\sin{\theta_{u/d}}, \nonumber \\
\langle H_{u/d}^{2} \rangle  &= \langle H_{u/d}^{SM}\rangle\sin{\phi_{u/d}} \sin{\theta_{u/d}}, \\
\langle H_{u/d}^{3} \rangle  &= \langle H_{u/d}^{SM}\rangle \cos{\theta_{u/d}}. \nonumber
\end{align}
Such Higgs VEVs can lead to flavor changing neutral currents (FCNC). We keep the mixing angles small and assume no problems with FCNC, which implies $\theta \ll 1$. 

\subsection{Toward Physical Coupling}\label{physical}
Note that the Yukawa couplings in M-theory belong to the high energy scale. We will attempt to use the already existent list of high scale Yukawa coupling running from SM experimental Yukawas in Table 1 of \cite{Babu:2016bmy} \footnote{The $GUT$ group is slightly different, but we assume the magnitude of the couplings are approximately the same. See also \cite{Ross:2007az}.} and find a solution for our parameters. We assume the effect of the extra U(1)'s from our theory in the renormalization group equations (RGEs) is not significant, and the Yukawas have approximately the same magnitudes as in \cite{Babu:2016bmy}.

In order to compare with physical Yukawa couplings, we need to take into account a few modifications. First, as mentioned in \cite{Beasley:2008kw}, we need an scaling factor to normalize the wave function. For cubic Yukawa, it is roughly proportional to $V_{G_2}^{-\frac{1}{2}}$ where $V_{G_2}$ is the volume of $G_2$ manifold and still a parameter in our theory (as local model cannot determine the global volume). Thus The scaling factor for all the cubic Yukawas is a parameter in this local model.
\subsection{Higgs VEVs}
One the other hand, recall that the Higgses only get VEVs at low scale. Therefore, precisely speaking, we can only consider the VEVs of the six Higgses after we run our M-theory Yukawa couplings down to low scale. Unfortunately, at high scale, we only have a set of algebraic expressions for M-theory Yukawas, making the running down to low scale complicated. Moreover, we cannot directly fit our Yukawas with the existing data of high scale running from SM Yukawas because they all assume a two Higgses model. Therefore, to remedy this problem, we will use a heuristic treatment assuming that the angular factors, in equations (\ref{spherical}), are regarded as part of the low scale Yukawa couplings and do not change much while running to high scale. Then, the effective VEVs at low scale are just the two VEVs from the SM, and the Yukawa couplings at high scale used to fit with Table 1 of \cite{Babu:2016bmy} then are
\begin{align}
Y = f(\phi, \theta) \lambda
\end{align}
where $\lambda$ is a Yukawa computed from section \ref{Yukawa_compute} and $f(\phi, \theta)$ is one of the angular functions associated with the Higgs fields from equations (\ref{spherical}.)  The full table of high scale Yukawa couplings with angular factors are presented in  Appendix \ref{Yukawatable}.

\section{Yukawa matrix for gauge group $SU(3)\times SU(2) \times U(1)^Y \times U(1)^b \times U(1)^c  \times U(1)^d$}\label{simplify}
First, we need to fix all extra degrees of freedom. Translation allows setting $v_d =0$. We also have three degrees of rotation and one degree of scaling to make $v_b =(1,0,0)$.

Second, we will try to consider the scattering around special cases of $H_b$ and $H_d$. Notice from the list in (\ref{cyclevolume}) that by setting all parameters to zero except $b$, we see that volumes of root $e_1-e_2$ and $e_2-e_3$ are controlled by $b$. They are responsible for breaking the adjoint of $E_8$ into three {\bf 27}'s of $E_6$ (see Figure \ref{breakingE8}), hence are also responsible for separating the three SM families.

On the other hand, $d$ controls $e_2-e_3$, $e_6-e_7$, and $e_7-e_8$. The blown-up two-cycle of $e_2-e_3$ breaks the adjoint of $E_8$ into two {\bf 27}'s of $E_6$, which transform as the fundamental and singlet of $SU(2)$ respectively, i.e, $({\bf 27}, \bf{2}) \oplus ({\bf 27},\bf{1})$. Thus $d$ seperates one family (the top quark family) from the other two in the adjoint of $E_8$. The latter still has an $SU(2)$ family symmetry (which is broken when we turn $b$ on ). Additionally, $e_6-e_7$ corresponds to breaking the {\bf 27}'s of $E_6$ into the presentations of $SO(10)$, separating the Higgses from quarks and leptons. Finally, $e_7-e_8$ splits the ${\bf 16}$'s of $SO(10)$ into the $\bf{10}$ and $\bar{\bf{5}}$ of $SU(5)$. Thus, $d$ also separates the up-type quarks (up, charm, top) from the down-type quarks (down, strange, bottom), i.e. an isospin breaking effect.
\section{Numerical Evaluation}\label{numerical}

To test the compatibility of this model with the Standard Model, we perform a regression on the free parameters by a least squares approach. Our calculations of Yukawa couplings are compared to experimentally measured weak scale Yukawa couplings which have been run up to the GUT scale \footnote{See also \cite{Ross:2007az}.}. The theoretical uncertainty in the calculation dominates over the experimental uncertainties and we only consider theoretical uncertainty when minimizing the sum of the residuals.

Using previous arguments, we set the base parameters corresponding to $a=0$  to zero, $v_d$ to zero, and $v_b$  to $(1,0,0)$. With three $3\times3$ traceless symmetric matrices $H_\phi$ and two $3-$vectors, we have 18 free parameters from the base space. We have four additional parameters from the Higgs VEVs, satisfying $\langle\big(H_1^2+H_2^2+H_3^2\big)^{1/2}\rangle=\langle H_{\text{MSSM}}\rangle$. Although we have more free parameters than constraints from the data, the non-linearity in calculating the Yukawas restricts the solutions. A list of numerical solutions is in Appendix.
\begin{figure}
	\begin{center}
		\includegraphics[width=90mm]{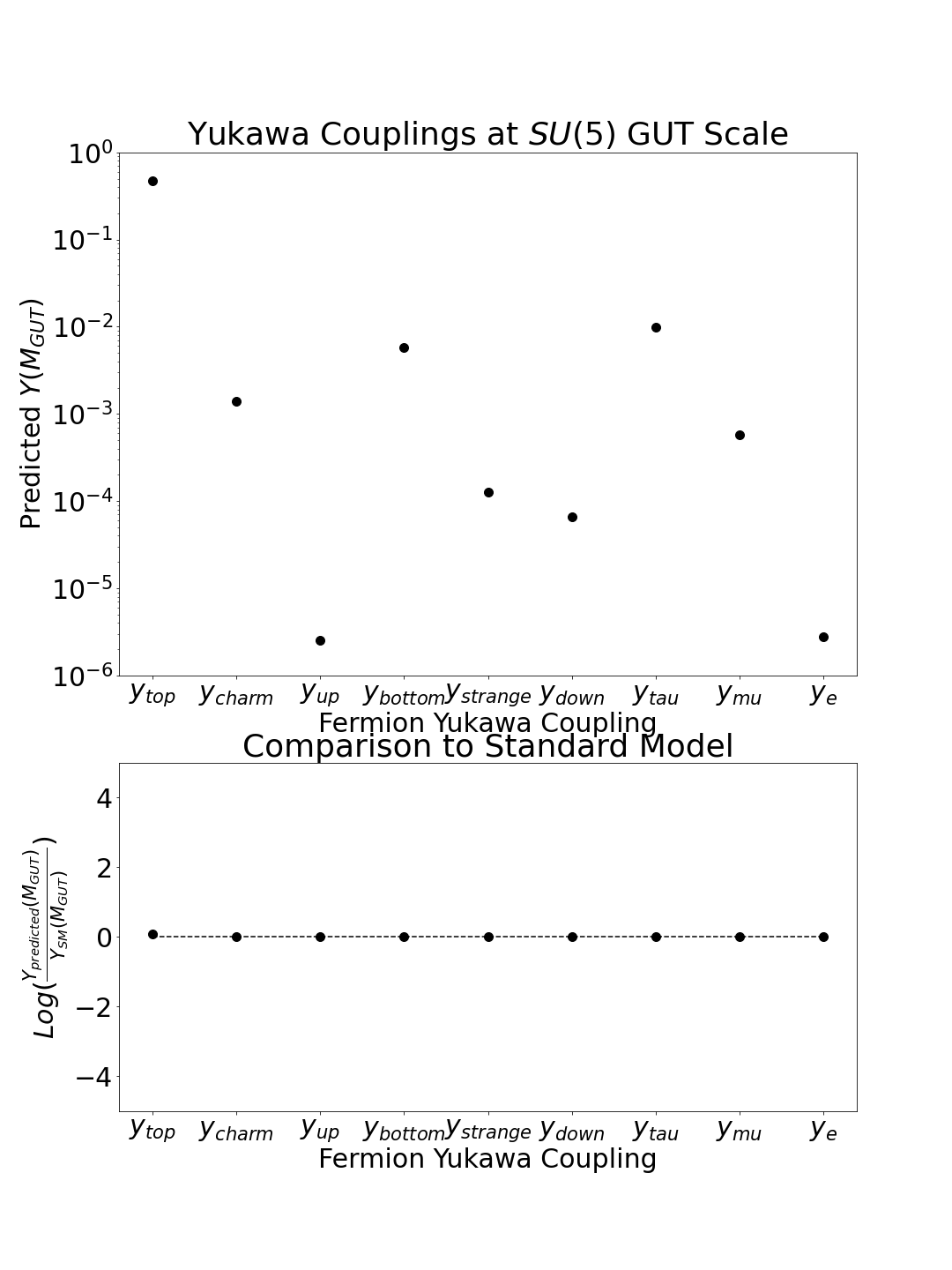}
		\caption{A set of sample solutions found numerically. The three symbols indicate three different solutions, and the line indicates the measured value for each Yukawa coupling.}
		\label{fig:num-samples}
	\end{center}
\end{figure}
A set of samples from numerical evaluation is shown in Fig.~\ref{fig:num-samples}. We have observed some general trends among the numerical solutions. Most importantly, there exists a hierarchy of Yukawas within each family which come from the breaking of the flavor and family symmetries. There is a large top quark Yukawa coupling. Finally, it appears that the hierarchy solution only happens when $\theta$ is small, an observation that is expected from the aforementioned no-neutral-current condition. 

\section{Effect of the Higgses and Yukawa couplings}\label{effect}
We want to use this section to emphasize the necessity of both the Higgs sector and the Yukawa exponential factor (which is of stringy origin) in satisfying the hierarchy. First, if only one family of the Higgses get VEVs, say $H_3$, we will get the up-type quark matrix of the form
\[\begin{Bmatrix}
&0 &A & 0\\
&A &0 &0\\
&0&0 &B
\end{Bmatrix}\].
Although we still have a hierarchy with one heavy and two light families. There is no hierarchy between the lighter two.

Second, if all three Higgs families get VEVs while all the Yukawa coefficients are the same (equal to 1), the theory will not have the physical hierarchy. Considering only the angular factors (dropping the common VEV factor), we have the matrix in the form
\[\begin{Bmatrix}
&0 &\tilde{A} & \tilde{B}\\
&\tilde{A} &0 &\tilde{C}\\
&\tilde{B} &\tilde{C} &\tilde{A}
\end{Bmatrix}\]
Then, from the characteristic equation, we conclude
\begin{equation}\label{constraint}
\left\{
\begin{array}{ll}
\lambda_1+\lambda_2+\lambda_3 =\tilde{A}\\
\lambda_1 \lambda_2 + \lambda_2 \lambda_3 + \lambda_3 \lambda_1 = \tilde{A}^2 + \tilde{B}^2 +\tilde{C}^2 =1	
\end{array}  
\right.
\end{equation}

This can be shown to imply that the quark hierarchy can never satisfy equations (\ref{constraint}). Therefore, both the three families of Higgses and the stringy Yukawa suppression are needed for the hierarchy. 
\section{Anomalies}\label{sec_anomaly}
The theory may result in  gauge boson triangle anomalies.  Such an anomaly can be canceled by St\"{u}ckelberg-Green-Schwarz mechanism and gives some bosons a mass.

\subsection{How to Compute the Anomaly}
We focus our attention on a model with gauge group $SU(3) \times SU(2) \times U(1)^n$ where the $U(1)$'s are to be examined. It can be shown that anomalies of the $U(1)$'s come from triangle loop of bosons in three configurations: $SU(3)-SU(3)-U(1)$ and $SU(2)-SU(2)-U(1)$ and $U(1)-U(1)-U(1)$. The anomaly of a triangle from three $U(1)$'s is proportional to the sum of particles  that transform under the nonabelian factor weighted by the charge of $U(1)$ factors. If this sum is zero, the configuration of $U(1)$'s is anomaly-free. Otherwise, it is anomalous.

Explicitly, for
\begin{itemize}
	\item $U^a(1)-U^b(1)-U^c(1)$ it is simply the sum, over all the particles, of the products of U(1) charges: $\sum_{i:\text{all particles}}q^a_i q^b_i q^c_i$.
	
	\item $SU(3)-SU(3)-U(1)$: Sum of U(1) charges over all triplet: $\sum_{i: \text{all triplets}} q_i$. 
	
	\item $SU(2)-SU(2)-U(1)$:  Sum of U(1) charges over all doublet: $\sum_{i: \text{all doublets}} q_i$. 
\end{itemize}

Note that $(\textbf{3}, \textbf{2})$ has three $SU(2)$ doublets and two $SU(3)$ triplets.

\begin{table}
	\caption{Anomaly computation.}\label{anomaly}
%
%
%
	
\end{table}

\subsection{Anomaly Cancelation by St\"{u}ckelberg-Green-Schwarz Mechanism}
\cite{Anastasopoulos:2006cz} In string theory, an additional term is added to cancel out the anomaly. Such a term will give a mass to the anomalous boson. This is called St\"{u}ckelberg-Green-Schwarz mechanism. The anomaly-related terms in effective action is
\begin{align}
\mathcal{S} =& -\sum_i \int d^4 x \frac{1}{4 g^2_i} F_{i, \mu \nu} F^{\mu \nu}_i - \underbrace{\frac{1}{2} \int d^4 x \sum_I (\partial_\mu a^I + M^I_i A^i_\mu)^2}_{\text{St\"{u}ckelberg term}}\\
&+ \underbrace{\frac{1}{24 \pi^2} C^I_{ij} \int a^I F^i \wedge F^j}_{\text{Green-Schwarz term}} + \underbrace{\frac{1}{24 \pi^2} E_{ij,k} \int A^i \wedge A^j \wedge F^k}_{\text{Chern-Simon term}}
\end{align}
where $a^I$ are axions, $C^I_{ij}$ is symmetric, $E_{ijk}$ is symmetric between $i$ and $j$.  Then, when the anomalous variation is distributed democratically among the three vertices, the condition for canceling the anomalies is

\begin{align}
t_{ijk} + E_{ijk} + E_{ikj} + M^I_i C^I_{jk}=0
\end{align}
where 	$t_{ijk}= \Tr \{t_i t_j t_k\}$. 
We now focus on the anomalies coming from $U(1)-U(1)-U(1)$ triangle which are computed in Table \ref{anomaly}. Then, the generators are commuting, so $t_{ijk}$ is totally symmetric. Summing all equations of permutation of $i, j,$ and $k$, we get
\begin{align}
M^I_i C^I_{jk} + M^I_j C^I_{ki} + M^I_k C^I_{ij} = -3 t_{ijk} \label{anomalyeq}
\end{align}

where we used $E_{ijk}= -E_{jik}$. We can use the value of $t_{ijk}$ to compute possible value for $M^I_i$ and $C^I_{ij}$. 

Notice that simultaneous transformation
\begin{align}
&	M^I_i \rightarrow a^I M^I_i &C^I_{ij} \rightarrow \frac{1}{a^I} C^I_{ij}
\end{align}
for all $i, j$ leaves the equations invariant. So, if (\ref{anomalyeq}) has a solution, the solution will only be unique up to the ratio of the masses. For the anomaly of $b-b-b$, the system is simply reduced to one linear equation giving $U(1)^b$ a nonzero mass, up to a scaling, 
\begin{align}
M_b = -3t_{bbb}= 18.
\end{align}
This specific number does not mean much due to scaling freedom \footnote{Study of anomaly involving $SU(2)$ and $SU(3)$ may fix this freedom.}. The only significant point is $U(1)^b$ being massive. Similarly, $U(1)^c$ is also massive. Unfortunately, $U(1)^d$ is anomaly-free and hence cannot get mass this way. Yet, as the Higgses are charged in $U(1)^d$ ($U(1)^b$ as well), their electroweak VEVs can give mass to the bosons.


\section{Conclusion}

In this paper, we use the geometric gauge breaking mechanism in M theory compactified on singular $G_2$ manifold to help understand quark and charged lepton masses.  We start with the adjoint representation of a single $E_8$ that contains exactly three related  families of quarks and leptons. Then, we break $E_8$ to the Standard Model via deformations and geometric engineering, following the technique of Katz and Morrison \cite{Katz:1996xe}. We explicitly computed Yukawa couplings in a local model and shows their fitting with experimental results.

With this approach, we hope to understand the origin of flavors and three families, and the values of quark and lepton masses.  We are partially successful.  We can see three families and the hierarchy of quark and lepton masses emerge. We can see the isospin breaking that makes the $SU(2)$ doublets such as top and bottom, up and down, electron and electron neutrino which all have different masses and the hierarchy of family masses.  The amounts are controlled by deformation parameters that are effectively moduli.  We can calculate the values of the deformation moduli that lead to the hierarchy and realistic values for the masses.  Ideally, we would be able to predict the values at which the deformation moduli are stabilized, and predict the masses, but we are not yet able to do so. In principal, the moduli have to satisfy stabilization constraints, neutrino sector, global $G_2$ structure, and so on. So, future study on these constraints applying to our quark and lepton context may make the theory predictive.

We are able to get some important mass values. We work with high scale Yukawa couplings.  The top quark has a Yukawa coupling of order one.  The up quark can be less than the down quark.  More precisely, $m_{up} + m_e \lesssim m_{down}$ (ignoring an electromagnetic contribution), so that protons will be stable rather than neutrons, allowing hydrogen atoms.  We can derive the conditions in the underlying theory for this inequality, or for the top Yukawa to be of order unity, but we cannot yet show they must uniquely hold.  Three families and a  hierarchy of masses do arise generically. The theory might not have allowed these results, so we view obtaining them in a UV complete theory as significant progress. We don’t at this stage have much control over what masses are associated with the three extra $U(1)$’s, but none should be massless. Then the spectrum should contain four new Z’ states.  They are well motivated. In future work it  may be possible to constrain their masses. Lastly, we also leave the study of the remaining particles resulted from $E_8$ breaking for future study. 
\appendix
\section*{Acknowledgments}
We would like to acknowledge the support from the LCTP at the University of Michigan and DoE grant DE-SC0007859. We also would like to thank all of Khoa's friends who helped editing  the manuscript.
\appendix
\section{Yukawa Tables}\label{Yukawatable}
Here, $n_{ij}$ takes value 1, -1, or 0 depending on the trivalent gradient flow existence and orientation whose details are in \cite{Braun:2018vhk}. We will assume they all 1 in this local model. $H$ and $v$ explicitly are
\begin{align}
&	H_{\phi}=
\begin{Bmatrix}
u_{\phi}^1 & u_{\phi}^3 & u_{\phi}^4\\
u_{\phi}^3 & u_{\phi}^2 & u_{\phi}^5\\
u_{\phi}^4 & u_{\phi}^5 & -u_{\phi}^1-u_{\phi}^2 \\ 
\end{Bmatrix},
&  v_{\phi} 
\begin{Bmatrix}
v_{\phi}^1\\
v_{\phi} ^2\\
v_{\phi} ^3
\end{Bmatrix}.
\end{align}
\begin{table}[H]
	\centering
	\caption {Up-type Quark terms.} 
	\scalebox{0.55}{\begin{tabular}{l |l}
		Term 	$Q_i u^c_j H^u_k$ & Coupling $Y^u_{ijk}$\\
		\hline
		$Q_1 u_2^c H^u_3 $ &  $n^u_{12}  \cos{\theta_{u}} \exp\Big\{$\\
		& $-\frac{1}{2} |(v_b-v_c  -v_d+v_Y)^T ( H_b  -H_c -H_d +H_Y)^{-1}( v_b -v_c-v_d+v_Y) +$\\
		&	$ ( - v_b-v_c -v_d-4v_Y)^T (- H_b  -H_c -H_d - 4H_Y)^{-1}( - v_b -v_c-v_d - 4v_Y) +$\\
		&	 $( 2v_d+3v_Y)^T ( 2H_c+2H_d +3H_Y)^{-1}(2v_d+3v_Y)| \Big\}$ \\ \hline
		$Q_1 u_3^c H^u_2$ & $n^u_{13} \sin{\phi_{u}} \sin{\theta_{u}}\exp\Big\{$\\
		& $-\frac{1}{2} |( v_b-v_c-v_d+v_Y)^T ( H_b  -H_c-H_d +H_Y)^{-1}( v_b-v_c-v_d+v_Y) +$\\
		&	$ (-v_c -v_d-4v_Y)^T (  -H_c-H_d - 4H_Y)^{-1}( -v_c -v_d - 4v_Y) +$\\ 
		&	 $( -v_b+2v_d+3v_Y)^T (-H_b  +2H_c+2H_d +3H_Y)^{-1}(-v_b +2v_d+3v_Y)| \Big\}$ \\ \hline		
		$Q_2 u_1^c H^u_3$ &   $n^u_{21} \cos{\theta_{u} } \exp\Big\{$\\
		& $-\frac{1}{2} |(-v_b-v_c -v_d+v_Y)^T ( -H_b -H_c -H_d +H_Y)^{-1}( -v_b -v_c-v_d+v_Y) +$\\
		&	$ ( v_b -v_c-v_d-4v_Y)^T ( H_b  -H_c -H_d - 4H_Y)^{-1}( v_b -v_c-v_d - 4v_Y) +$\\
		&	 $(  2v_d+3v_Y)^T (2H_c+2H_d +3H_Y)^{-1}(2v_d+3v_Y)| \Big\}$ \\ \hline
		$Q_2 u_3^c H^u_1$ & $n^u_{23}  \cos{\phi_{u}}\sin{\theta_{u}}\exp\Big\{$\\
		& $-\frac{1}{2} |(-v_b -v_c-v_d+v_Y)^T ( -H_b  -H_c-H_d +H_Y)^{-1}( -v_b-v_c -v_d+v_Y) +$ \\
		&	$ ( -v_c-v_d-4v_Y)^T (  -H_c -H_d - 4H_Y)^{-1}(  -v_c-v_d - 4v_Y) +$\\
		&	 $( v_b+2v_c+2v_d+3v_Y)^T (H_b  +2H_c+2H_d +3H_Y)^{-1}(v_b+2v_c +2v_d+3v_Y)| \Big\}$ \\ \hline
		$Q_3 u_1^c H^u_2$ &  $n^u_{31}\sin{\phi_{u}} \sin{\theta_{u}} \exp\Big\{$\\
		& $-\frac{1}{2} |(-v_c -v_d+v_Y)^T (  -H_c -H_d +H_Y)^{-1}(-v_c -v_d+v_Y) +$\\
		&	$ (v_b-v_c -v_d-4v_Y)^T ( H_b -H_c -H_d - 4H_Y)^{-1}( v_b -v_c-v_d - 4v_Y) +$\\ 
		&	 $( -v_b+2v_c+2v_d+3v_Y)^T (-H_b  +2H_c+2H_d +3H_Y)^{-1}(-v_b 2v_c+2v_d+3v_Y)| \Big\}$ \\ \hline
		$Q_3 u_2^c H^u_1$ &  $n^u_{23}\cos{\phi_{u}}\sin{\theta_{u}} \exp\Big\{$\\
		& $-\frac{1}{2} |( -v_c-v_d+v_Y)^T (   -H_c -H_d +H_Y)^{-1}(-v_c -v_d+v_Y) +$ \\
		&	$ (-v_b-v_c -v_d-4v_Y)^T ( -H_b  -H_c -H_d - 4H_Y)^{-1}( -v_b -v_c-v_d - 4v_Y) +$\\
		&	 $( v_b+2v_c+2v_d+3v_Y)^T (H_b +2H_c+2H_d +3H_Y)^{-1}(v_b +2v_c+2v_d+3v_Y)| \Big\}$ \\ \hline
		$Q_3 u_3^c H^u_3$ &  $ n^u_{33}  \cos{\theta_{u} } \exp\Big\{ $\\
		& $-\frac{1}{2} |(-v_c -v_d+v_Y)^T (  -H_c-H_d +H_Y)^{-1}( -v_c -v_d+v_Y) +$\\
		&	$ (-v_c -v_d-4v_Y)^T (  -H_c -H_d - 4H_Y)^{-1}( -v_c -v_d - 4v_Y) +$\\
		&	 $(2v_c+ 2v_d+3v_Y)^T (2H_c+2H_d +3H_Y)^{-1}(2v_c+2v_d+3v_Y)| \Big\}$ \\ \hline
		All else & 0\\ \hline
	\end{tabular}}	
\end{table} 

\begin{table}[H]
	\centering
	\caption {Down-type Quark terms.} \label{tab:title}
	\scalebox{0.55}{\begin{tabular}{@{} l |l @{}}
		\hline
		Term 	$Q_i d^c_j H^d_k$ & Coupling $Y^d_{ijk}$\\
		\hline
		
		$Q_1 d_2^c H^d_3$ & $n^d_{12}\cos{\theta_{d}}  \exp\Big\{$\\
		&$-\frac{1}{2} |(v_b-v_c -v_d+v_Y)^T ( H_b -H_c -H_d +H_Y)^{-1}( v_b-v_c -v_d+v_Y) +$\\
		&	$ ( - v_b-v_c +3v_d+2v_Y)^T (- H_b-H_c +3H_d +2H_Y)^{-1}( - v_b -v_c+3v_d +2v_Y) +$\\
		&	 $(  2v_c-2v_d-3v_Y)^T (2H_c -2H_d -3H_Y)^{-1}(2v_c-2v_d-3v_Y)| \Big\}$  \\ \hline
		$Q_1 d_3^c H^d_2$ & $n^d_{13}  \sin{\phi_{d}}\sin{\theta_{d}} \exp\Big\{$\\
		&$-\frac{1}{2} |( v_b-v_c-v_d+v_Y)^T ( H_b -H_c-H_d +H_Y)^{-1}( v_b-v_c-v_d+v_Y) +$\\
		&	$ (-v_c+ 3v_d+2v_Y)^T (-H_c+ 3H_d +2H_Y)^{-1}(-v_c+  3v_d +2v_Y) +$\\ 
		&	 $(2v_c -v_b-2v_d-3v_Y)^T (-H_b +2H_c -2H_d -3H_Y)^{-1}(2v_c-v_b -2v_d-3v_Y)| \Big\}$ \\ \hline
		$Q_2 d_1^c H^d_3$ &    $n^d_{21} \cos{\theta_{d}} \exp\Big\{$\\
		&$-\frac{1}{2} |(-v_b -v_c-v_d+v_Y)^T ( -H_b -H_c -H_d +H_Y)^{-1}( -v_b-v_c -v_d+v_Y) +$\\
		&	$ ( v_b-v_c +3v_d+2v_Y)^T ( H_b -H_c +3H_d +2H_Y)^{-1}( v_b -v_c+3v_d +2v_Y) +$\\
		&	 $(2v_c -2v_d-3v_Y)^T (2H_c-2H_d -3H_Y)^{-1}(2v_c-2v_d-3v_Y)| \Big\}$ \\ \hline
		$Q_2 d_3^c H^d_1$& $n^d_{23}  \cos{\phi_{d}}\sin{\theta_{d}}\exp\Big\{$\\
		&$-\frac{1}{2} |(-v_b -v_c-v_d+v_Y)^T ( -H_b -H_c -H_d +H_Y)^{-1}( -v_b-v_c -v_d+v_Y) +$ \\
		&	$ ( -v_c+3v_d+2v_Y)^T ( -H_c+3H_d +2H_Y)^{-1}( -v_c+3v_d + 2v_Y) +$\\
		&	 $( v_b+2v_c-2v_d-3v_Y)^T (H_b+2H_c  -2H_d -3H_Y)^{-1}(v_b+2v_c -2v_d-3v_Y)| \Big\}$ \\ \hline
		$Q_3 d_1^c H^d_2$ &  $n^d_{31} \sin{\phi_{d}}\sin{\theta_{d}}\exp\Big\{$\\
		&$-\frac{1}{2} |( -v_c-v_d+v_Y)^T ( -H_c-H_d +H_Y)^{-1}(-v_c -v_d+v_Y) +$\\
		&	$ (v_b-v_c +3v_d+2v_Y)^T ( H_b -H_c+3H_d + 2H_Y)^{-1}( v_b -v_c+3v_d +2v_Y) +$\\ 
		&	 $( -v_b+2v_c-2v_d-3v_Y)^T (-H_b +2H_c -2H_d -3H_Y)^{-1}(-v_b +2v_c-2v_d-3v_Y)| \Big\}$ \\ \hline
		$Q_3 d_2^c H^d_1$ &  $n^d_{32} \cos{\phi_{d}}\sin{\theta_{d}} \exp\Big\{ $\\
		&$-\frac{1}{2} |( -v_c-v_d+v_Y)^T ( -H_c -H_d +H_Y)^{-1}( -v_c-v_d+v_Y) +$ \\
		&	$ (-v_b-v_c +3v_d+2v_Y)^T ( -H_b  -H_c+3H_d +2H_Y)^{-1}( -v_b-v_c +3v_d +2v_Y) +$\\
		&	 $( v_b+2v_c-2v_d-3v_Y)^T (H_b +2H_c -2H_d -3H_Y)^{-1}(v_b+2v_c -2v_d-3v_Y)| \Big\}$ \\ \hline
		$Q_3 d_3^c H^d_3$ & $ n^d_{33}\cos{\theta_{d}} \exp\Big\{$\\
		&$-\frac{1}{2} |(-v_c-v_d+v_Y)^T ( -H_c-H_d +H_Y)^{-1}(  -v_d+v_Y) +$\\
		&	$ ( -v_c+3v_d+2v_Y)^T (-H_c +3H_d +2H_Y)^{-1}(  -v_c+3v_d +2v_Y) +$\\
		&	 $(2v_c-2v_d-3v_Y)^T (2H_c -2H_d -3H_Y)^{-1}(2v_c-2v_d-3v_Y)| \Big\}$  \\ \hline
		All else & 0 \\ \hline
	\end{tabular}}
\end{table}

\begin{table}[H]
	\centering
	\caption {Electron-type terms.} \label{electron}
	\scalebox{0.55}{\begin{tabular}{@{} l |l @{}}
		\hline
		Term 	$L_i e^c_j H^d_k$ & Coupling $Y^l_{ijk}$\\
		\hline
		$L_1 e^c_2 H^d_3$ &  $n^e_{12}  \cos{\theta_{d}}\exp\Big\{$\\
		&$-\frac{1}{2} |(v_b  -v_c+3v_d-3v_Y)^T ( H_b  -H_c+3H_d -3H_Y)^{-1}( v_b-v_c +3v_d-3v_Y) +$\\
		&	$ (- v_b -v_c-v_d+6v_Y)^T (-H_b -H_c -H_d+6H_Y )^{-1}( -v_b -v_c-v_d+6v_Y) +$\\
		&	 $(2v_c -2v_d-3v_Y)^T ( 2H_c-2H_d -3H_Y)^{-1}(2v_c-2v_d-3v_Y)| \Big\}$ \\ \hline
		$L_1 e^c_3 H^d_2$ & $n^e_{13} \sin{\phi_{d}}\sin{\theta_{d}} \exp\Big\{ -\frac{1}{2} |(v_b  -v_c+3v_d-3v_Y)^T ( H_b  -H_c+3H_d -3H_Y)^{-1}( v_b -v_c+3v_d-3v_Y) +$\\
		&	$ ( -v_c -v_d+6v_Y)^T ( -H_c-H_d+6H_Y)^{-1}( -v_c -v_d+6v_Y) +$\\ 
		&	 $( -v_b+2v_c-2v_d-3v_Y)^T (-H_b +2H_c -2H_d -3H_Y)^{-1}(-v_b+2v_c -2v_d-3v_Y)| \Big\}$ \\ \hline
		
		$L_2 e^c_1 H^d_3$ &   $n^e_{21} \cos{\theta_{d}}\exp\Big\{$\\
		&$-\frac{1}{2} |(-v_b -v_c +3v_d-3v_Y)^T (- H_b -H_c +3H_d -3H_Y)^{-1}(- v_b -v_c+3v_d-3v_Y) +$\\
		&	$ ( v_b-v_c -v_d+6v_Y)^T ( H_b -H_c -H_d +6H_Y)^{-1}( v_b  -v_c-v_d+6v_Y) +$\\
		&	 $(2v_c - 2v_d-3v_Y)^T (2H_c-2H_d -3H_Y)^{-1}(2v_c-2v_d-3v_Y)| \Big\}$ \\ \hline
		$L_2 e^c_3 H^d_1$ & $n^e_{23}  \cos{\phi_{d}}\sin{\theta_{d}}\exp\Big\{$\\
		&$-\frac{1}{2} |(-v_b  -v_c+3v_d-3v_Y)^T (- H_b -H_c +3H_d -3H_Y)^{-1}(- v_b-v_c +3v_d-3v_Y)  +$ \\
		&	$( -v_c -v_d+6v_Y)^T ( -H_c-H_d+6H_Y)^{-1}(  -v_c-v_d+6v_Y) +$\\
		&	 $( v_b+2v_c-2v_d-3v_Y)^T (H_b +2H_c -2H_d -3H_Y)^{-1}(v_b+2v_c -2v_d-3v_Y)| \Big\}$ \\ \hline
		$L_3 e^c_1 H^d_2$ &  $n^e_{31} \sin{\phi_{d}}\sin{\theta_{d}} \exp\Big\{$\\
		&$-\frac{1}{2} |(-v_c + 3v_d-3v_Y)^T ( -H_c+3H_d -3H_Y)^{-1}(-v_c+3v_d-3v_Y) +$\\
		&	$ (v_b  -v_c-v_d+6v_Y)^T ( H_b  -H_c-H_d+6H_Y)^{-1}( v_b-v_c  -v_d+6v_Y) +$\\ 
		&	 $( -v_b+2v_c-2v_d-3v_Y)^T (-H_b+2H_c  -2H_d -3H_Y)^{-1}(-v_b+2v_c -2v_d-3v_Y)| \Big\}$ \\ \hline
		$L_3 e^c_2 H^d_1$ &  $n^e_{32} \cos{\phi_{d}}\sin{\theta_{d}}\exp\Big\{$\\
		&$-\frac{1}{2} |( -v_c+3v_d-3v_Y)^T ( -H_c+3H_d -3H_Y)^{-1}( -v_c+3v_d-3v_Y) +$ \\
		&	$  (- v_b-v_c  -v_d+6v_Y)^T (-H_b -H_c -H_d+H_Y )^{-1}( -v_b -v_c -v_d+6v_Y) +$\\
		&	 $( v_b+2v_c-2v_d-3v_Y)^T (H_b +2H_c-2H_d -3H_Y)^{-1}(v_b+2v_c -2v_d-3v_Y)| \Big\}$ \\ \hline
		$L_3 e^c_3 H^d_3$ &  $ n^e_{33} \cos{\theta_{d}} \exp\Big\{$\\
		&$-\frac{1}{2} |(-v_c+ 3v_d-3v_Y)^T ( -H_c+3H_d -3H_Y)^{-1}(-v_c+ 3v_d-3v_Y) +$\\
		&	$( -v_c -v_d+6v_Y)^T ( -H_c-H_d+6H_Y)^{-1}(  -v_c -v_d+6v_Y ) +$\\
		&	 $( 2v_c-2v_d-3v_Y)^T (2H_c-2H_d -3H_Y)^{-1}(2v_c-2v_d-3v_Y)| \Big\}$ \\ \hline
		All else & 0 \\ \hline
	\end{tabular}}	
\end{table}

The numerical result for the moduli used in Fig \ref{fig:num-samples} is in table \ref{num-table2}

\begin{table}[H]
	\centering
	\caption {Tabulated numerical values of Moduli} \label{num-table2}
	\scalebox{0.55}{\begin{tabular}{c | c || c | c}
		\hline
        Parameter & Value & Parameter & Value \\
        \hline \hline
        $H_{b \ 1, 1}$ & $1.75219628381272$   & $H_{Y \ 1, 1}$ & $0.470254977486118$ \\
        $H_{b \ 1, 2}$ & $-0.735328652705781$ & $H_{Y \ 1, 2}$ & $0.701824648617083$ \\
        $H_{b \ 1, 3}$ & $-0.377020719433746$ & $H_{Y \ 1, 3}$ & $-1.34973735409641$ \\
        $H_{b \ 2, 2}$ & $1.19315995302413$   & $H_{Y \ 2, 2}$ & $0.641604847709697$ \\
        $H_{b \ 2, 3}$ & $0.675543994721913$  & $H_{Y \ 2, 3}$ & $0.108762493856499$ \\
        $H_{c \ 1, 1}$ & $1.41562893856031$   & $v_{c \ 1}$ & $-0.250519055696569$ \\
        $H_{c \ 1, 2}$ & $1.05931181064608$   & $v_{c \ 2}$ & $-1.4656911323334$ \\
        $H_{c \ 1, 3}$ & $1.45371284442019$   & $v_{c \ 3}$ & $-0.405862359379641$ \\
        $H_{c \ 2, 2}$ & $0.509312354295455$  & $v_{Y \ 1}$ & $-0.417353540440179$ \\
        $H_{c \ 2, 3}$ & $-1.11688627765089$  & $v_{Y \ 2}$ & $0.688596965400472$ \\
        $H_{d \ 1, 1}$ & $-1.13434995566124$  & $v_{Y \ 3}$ & $-1.50894939341168$ \\
        $H_{d \ 1, 2}$ & $-0.386481794947734$ & ${\phi}_1$   & $0.439157499240369$ \\
        $H_{d \ 1, 3}$ & $-1.34973735409641$  & ${\theta}_1$ & $-1.37741142245986$ \\
        $H_{d \ 2, 2}$ & $0.336495907974189$  & ${\phi}_2$   & $3.76691301205782$ \\
        $H_{d \ 2, 3}$ & $2.25852327647213$   & ${\theta}_2$ & $6.06985994999668$ \\
        \hline
	\end{tabular}}
\end{table}

\bibliography{citations}

\end{document}